\documentclass[12pt]{article}
\usepackage{amsmath,amssymb,graphicx,mathrsfs,hyperref,slashed,geometry}
\newcommand{\be}{\begin{equation}}
\newcommand{\ee}{\end{equation}}
\newcommand{\bea}{\begin{eqnarray}}
\newcommand{\eea}{\end{eqnarray}}

\def\({\left(} \def\){\right)}

\renewcommand{\baselinestretch}{1.25}
\begin{document}
\title{\vspace{-1.8in}
{A maximal-entropy initial state of the Universe  \\ as a microscopic description of inflation }}

\author{\large Ram Brustein${}^{(1,2)}$, A.J.M. Medved${}^{(3,4)}$
\\
\vspace{-.5in} \hspace{-.05in}  \vbox{
\begin{flushleft}
$^{\textrm{\normalsize
(1)\ Department of Physics, Ben-Gurion University,
Beer-Sheva 84105, Israel}}$
$^{\textrm{\normalsize
(2)\ Theoretical Physics Department, CERN, 1211 Geneva 23, Switzerland}}$
$^{\textrm{\normalsize (3)\ Department of Physics \& Electronics, Rhodes University,
Grahamstown 6140, South Africa}}$
$^{\textrm{\normalsize (4)\ National Institute for Theoretical Physics (NITheP), Western Cape 7602,
South Africa}}$
\\ \small \hspace{1.07in}
ramyb@bgu.ac.il,\ j.medved@ru.ac.za
\end{flushleft}
}}
\date{}
\maketitle
\begin{abstract}
We propose that the initial state of the Universe was an isotropic state of maximal entropy. Such a state can be described in terms of a state of closed, interacting, fundamental strings in their high-temperature Hagedorn phase. The entropy density in this state is equal to the square root of the energy density in Planck units, while the pressure is positive and equal to the energy density. These relations imply a maximally large entropy density and, therefore, a state that cannot be described by a semiclassical spacetime geometry. If one nevertheless insists on an effective semiclassical description of this state, she can do so by ignoring the entropy.  This leads  to a partially equivalent description in which the pressure appears to be negative and equal in magnitude to the energy density, as if the energy-momentum tensor was that of a cosmological constant. From this effective perspective, the state describes a period of string-scale inflation. The bound state of strings ultimately decays, possibly by a process akin to Hawking radiation, and  undergoes a transition into a phase of hot radiation. But, from the effective perspective, the same decay corresponds to the heating of the Universe at the end of inflation. Small quantum mechanical fluctuations in the initial state lead to a scale-invariant temperature anisotropies in the hot radiation. The temperature anisotropies are interpreted in the effective description as arising from quantum fluctuations of the curvature and an effective inflaton field. The stringy microscopic description determines the parameters of the model of inflation, as well as the cosmological observables, in terms of the string length scale and coupling strength. Our framework is similar, conceptually, to a recent description of black holes in terms of a maximal entropy state of strings in the Hagedorn phase.
\end{abstract}
\renewcommand{\baselinestretch}{1.5}\normalsize

\section{Introduction}

The hot big-bang model provides an accurate description of the cosmological evolution of our Universe starting from a thermal state of hot  radiation. However, the same model also faces some unresolved issues that are related to properties of the initial state: the large-scale smoothness,  the small-scale inhomogeneities and the smallness of the spatial curvature.   As a way to extend the validity of the hot big-bang model and resolve some of its shortcomings, it was proposed that, prior to the thermal phase, the Universe expanded exponentially during a relatively long period of cosmic inflation. The Universe then ``reheated" in an event that marks the beginning of the thermal phase; this being the essence of  the inflationary paradigm \cite{lindebook,guthbook}.

This  paradigm provides a framework for an effective description of the period of  exponential expansion. The scale of inflation, its duration and the type of matter that drives it ---  the so-called inflaton field ---  are all undetermined and subject to rather weak constraints. For any given  model of inflation, these parameters need to be specified in a way that  predicts the cosmological observables; in particular, the spectral  properties of the observed temperature inhomogeneities.  Many of the puzzles surrounding the inflationary paradigm are associated with attempts to identify the inflaton field with one of the fields in the standard model or in some type of grand unified theory or in some version of string theory.

The inflationary paradigm has been criticized recently \cite{steinh,brandnew} (see also \cite{martin}) on the grounds that a weakly coupled  model of inflation is not self-consistent when implemented within semiclassical gravity. More specifically,  the  self-reproduction property of such models leads to
scenarios of eternal inflation which imply that semiclassical evolution breaks down.  Furthermore, the scale of inflation that is needed to reproduce the correct spectra of perturbations  must be much lower than the natural cutoff of the theory, the Planck scale.

Meanwhile,  the problem of an initial singularity  continues to persist in the inflationary paradigm. This concern is well known \cite{Borde1,Borde2} but sometimes dismissed by arguing that the  singularity decouples from horizon-scale physics.
The premise being that  cosmological observables, which are determined solely by the latter, will not be affected by whatever does resolve the singularity. However, this is not a truly  admissible argument because,  in the words of Hawking
\cite{Hawking:1994ss},
 ``The only way to have a scientific theory is if the laws of physics hold everywhere including at the beginning of the Universe.''  Some remnants of the problem linger and reappear when one attempts to define a measure on the space of initial conditions. Just like  the analogous problem  for a  black hole (BH) \cite{bookdill}, it turns out that the resolution is much more surprising than a small tweak  which  transforms  the singularity into an epoch of large but finite curvature.

Here, we propose a microscopic model for the  state of the Universe when it is at the highest sustainable temperature for a state of  strings, the Hagedorn temperature. This state replaces the past of the hot big-bang Universe, resolves the singularity and provides the initial conditions for the subsequent  evolution of the  thermal radiation and the semiclassical cosmological geometry.

Our microscopic model is guided, in large part,  by the polymer model of BHs \cite{strungout,emerge}, which  suggests that a maximally large entropy  is an essential feature of non-singular gravitational states.  Maximal entropy in this context means the  saturation of the causal entropy bound \cite{ceb,hubblebound}; then  the entropy density $s$ is as large as it can be in relation  to  the energy density $\rho$ in appropriate units. Fundamental,  closed strings in the Hagedorn phase~\footnote{The current  proposal differs from previous ones that involved the Hagedorn phase of strings, such as string-gas cosmology \cite{brandgas1,brandgas2}.} are in just such  a  maximally entropic state.  Just like for BHs, the microscopic description of cosmology in  terms of this  hot string  state is non-singular.  In both cases, the apparent singularity is resolved  by making  dramatic  changes to the  state on horizon-sized scales. However, whereas the changes for a BH  are to its interior region, it is  the pre-history of the thermal state that is changed in the cosmological picture. Whether one is talking about BHs or  cosmology,  the cost of regularization  is that the hot string state cannot be described by  a semiclassical geometry.

Even if lacking a geometry,  a state of hot, interacting, closed  strings can, as  discussed in \cite{strungout}, be described by a simple free energy which is expressible as a power series in  the entropy density. For the BH polymer, $s$ and the other thermodynamic densities had to have non-trivial radial profiles. In the context of cosmology, however,  approximate isotropy and homogeneity are now mandatory features;  meaning that  the free energy  and all of its associated  thermodynamic densities are
approximately constant in space and time.

Although the current proposal is conceptually similar to our proposed resolution of the BH singularity \cite{bookdill}, it does differ in some important ways that go beyond questions about spacetime (in)dependence. For instance, who plays the role of the ``score-keeping  observer''? In the case of BHs, it is clear that the asymptotic external observer serves this purpose.  The cosmological analogue --- perhaps not quite as obvious ---   is the late-time   or  ``Friedmann--Robertson--Walker (FRW) observer''.  This is because  the past of this observer, before the beginning of the hot-radiation phase, is the analogue of the interior of the BH. As similarly  argued for the BH case in \cite{BHfollies}, all   proposals for the pre-history are perfectly acceptable as long as they are  self-consistent, able to reproduce the  observable Universe and compatible with the laws of physics (see also \cite{nima}). From the microscopic point of view, the puzzles of the FRW observer originate from trying to explain the initial quantum state using  effective semiclassical terms. The same situation was prevalent for BHs  and led to their infamous paradoxes. As will be shown here, the FRW observer can (and usually does) interpret our  maximally entropic state as one of vanishing entropy (and/or temperature) and  approximately describes it by using the flat-space slicing of a de Sitter (dS) spacetime.

In our cosmological model, a  geometric description of the past may be absent, but one can still adopt the equivalent  representation of gravity as an inertial effect in a conformally flat space.~\footnote{The choice of conformally flat rather than just flat  is to accommodate the constancy of the thermodynamic densities as discussed above. This point is elaborated on in Subsection~2.1.} Conformally flat spacetime coordinates, $t_{st}$, $r_{st}$, {\em etc.}, would then  represent labels for the position of the strings but physical observables will  not depend on these fiducial coordinates.  From this point of view, gravity is an emergent effect; it is a long-distance description of the microscopic forces between  constituents.  It is only when  gravity is semiclassical
that both of the descriptions,  geometric and inertial,  can co-exist.

In the microscopic model, the scale of inflation is fixed by the string scale
and the duration of  the exponential expansion, as perceived by the FRW observer, is the logarithm of the entropy of the Universe in natural units. In this sense, inflation is as short as it can be.
But the maximally entropic state still  needs to be large enough  to describe a large Universe, so that one could also try to understand how the initial state came to be so large. We will defer this issue to a future investigation but still want to suggest some possibilities.  For example, the Universe could start out in  a large, weakly curved, contracting phase for which it  was initially large and empty ---  as in  pre-big-bang scenarios \cite{pbb} or ekpyrotic models
\cite{ekpyrotic} ---  and then undergo a phase of contraction in which the number of Hubble-sized regions grows and the strings eventually heat up to the Hagedorn temperature. This would be the analogue of a large matter system  collapsing to form
a BH  \cite{DV}.  However, it is not important for current purposes to track the pre-history of the initial state. A good choice of an initial state is always a part of a good description of any physical system.

Throughout the paper,  we typically  focus on the case of $\;D=d+1=3+1\;$ spacetime dimensions. However, our conclusions are unchanged for  $\;D>4\;$.

\section{Initial state of the Universe}

\subsection{Microscopic perspective: Hagedorn phase of fundamental strings}

Closed strings in the Hagedorn phase have an exponentially large density of states
({\em e.g.}, \cite{FV,AW,Deo}) and are, therefore,  particularly well suited for describing states with high entropy.  Moreover, the equation of state connecting the pressure $p$ and the energy density $\rho$ for such high-temperature string theory is known to be $\;p=\rho\;$.  And so  $p$ is as large as a single-fluid pressure could be while still respecting causality. The standard thermodynamic relation  $\;\rho+p=sT\;$   for the energy density, pressure,   entropy density $s$ and temperature $T$  then leads to     $\;s=\sqrt{\rho}\;$
in Planck units, from which one obtains  $\;1/T=ds/d\rho=s/2\rho\;$  or $\;sT=2\rho\;$.  Meaning that $s$ is also as large as it could ever be in comparison to $\rho$.

An important consequence of the maximal entropy and pressure is that these require  exact spatial flatness and the vanishing of the  cosmological constant. This is because the  introduction of any such  sources would reduce the ratio $p/\rho$ away from unity.

We now  briefly review, for completeness,  relevant parts of the  discussion in \cite{strungout,emerge} and then adapt them to the current cosmological setup.

Both the energy $E$ and entropy $S$ of non-interacting, closed  strings in the Hagedorn phase scale linearly with the total length of string $\;L=Nl_s\;$,
so that  $\;S\sim N\;$ and $\;E\sim N/l_s\;$ ($l_s$ is the string length scale).
The free energy  $\;F=E-ST\;$  vanishes for non-interacting strings at the Hagedorn temperature $\;T_{Hag}= M_s/(4\pi)\;$ ($M_s=1/l_s\;$ is the string mass).   Then, for temperatures $T$ close to $T_{Hag}$, the free energy $F$  should be parametrically smaller than $N/l_s$.  We will use a  dimensionless parameter $\epsilon$ to  parametrize  this small  number,  $\;\epsilon=(T-T_{Hag}) /T_{Hag}\ll 1\;$. For the case of interacting strings, it makes sense to consider both positive and negative $\epsilon$; however, one can expect a phase transition when $\;\epsilon< 0\;$ (see below).

Interactions between strings take place at their intersections. The strength of the interactions --- that is,  the probability that two different strings join to form  one single longer loop or that a single string splits up into two shorter loops ---   is determined by the string coupling $g$ and is equal to $g^2$. For weakly coupled strings, $\;g^2 = M_s^2/M_P^2\;$, where $M_P$ is the Planck mass.  If $V$ is the volume of the region of space that is occupied by the strings (in terms of flat fiducial coordinates as explained in Section~1),
then  the volume density of intersections is   $N^2/V$, and  so  the leading-order effects of  interactions  can be described by including  a term scaling as
$\;g^2 N^2/V=N(g^2N/V)\;$ in the free energy.  Hence,
the free energy for interacting strings in the Hagedorn phase is  expressible  in terms of the entropy density $\;s = N/V\;$   as follows
\cite{strungout,emerge}:
\be
-\left(\frac{F}{V T_{Hag}}\right)_{strings}\;= \;\epsilon s - \frac{1}{2} g^2  s^2 + \cdots\;,
\label{FES2}
\ee
where string units ($l_s=1$) have now been adopted and  the first term on the right follows directly from previous relations
given that  $\;\rho=E/V=N T_{Hag}/V\;$. The ellipsis denotes  higher-order interactions, which  are small under the conditions that we consider, to be discussed in more detail later. Not coincidentally, this free energy is formally similar to those in the literature on  interacting polymers ({\em e.g.}, \cite{polytext}).

Extremizing the free energy with respect to $s$, one finds that
\be
\label{consol}
s \;=\; \frac{\epsilon}{g^2}\;.
\ee
The  energy density and pressure now follow from   the standard thermodynamic definitions,
\bea
\label{rhostring}
\rho &=&  \frac{1}{2}\frac{\epsilon^2}{g^2}\;,
 \\
 \label{pstring}
p&=&  \frac{1}{2}\frac{\epsilon^2}{g^2}\;,
\eea
and so the equation of state of the associated fluid is indeed $\;p=\rho\;$.  Comparing Eqs.~(\ref{consol}) and (\ref{rhostring}), one obtains the similarly advertised form
$\;s= 2 \sqrt{\rho/g^2}\;$ or, more simply,  $\;s\epsilon=2\rho\;$. The latter
suggests that $\epsilon$ acts like an effective temperature,  not to be confused with the local temperature of the string state which is  the Hagedorn temperature. Whereas the BH version of  $\epsilon$ turned out to be the Hawking temperature, it will be (re)calculated later on  and identified as the Gibbons--Hawking temperature \cite{GH}.

Let us recall that, for the  BH scenario, $\epsilon$ and thus the various
densities have  radial profiles.~\footnote{This dependence was not always made explicit
in previous papers.}  As the cosmological fluid has to be approximately
isotropic and homogeneous, the densities $s$ and $\rho$, as well as the pressure $p$ and the effective temperature $\epsilon$ will  all be regarded as  constants. Although, near the boundary of the string state, there is likely to be  some deviations from this constancy.

To specify the solution completely, $\epsilon$ needs to be fixed in a way that is
independent of Eq.~(\ref{consol}). For this, we will relate $\epsilon$ to
 the causal connection scale $R_{CC}$ \cite{ceb},  which can,  in general,  be viewed as the relativistic analogue of the Jeans length.  The notion of a Jeans length is applicable in the current context because of the fact that, at long distances, the string interactions are dominated by gravity \cite{LT,HP,DV}. For BHs, $R_{CC}$ corresponds to  the Schwarzschild radius and
thus serves as the linear scale for the whole string state. But the cosmological string state can encompass many causally disconnected regions; meaning  that $R_{CC}$ must be the linear scale as measured in fiducial coordinates of
just one such region.

There are several ways in which $R_{CC}$ can be estimated.  It was originally estimated in \cite{ceb} in its capacity as a ``Jeans-like" length scale, which means considering the equation of motion for a generic perturbation.  Because the string state is translation invariant, it is
 convenient
if  perturbations are expressed in their Fourier-space form $\delta_k$;
in which case, the relevant equation is
\be
\ddot{\delta}_k + \left(k^2-{R^{-2}_{CC}}\right) \delta_k \;=\;0\;,
\label{perteq}
\ee
where $\;k=|\vec{k}|\;$ is the wavenumber and  a dot represents a time derivative, both  with respect to the fiducial coordinates.
Following \cite{ceb}, one will find that
\be
R_{CC}^{-2} \;\simeq\; g^2~\text{Max} [ \rho/3 - p , \rho + p]\;.
\ee
When $\;p=0\;$,  the causal connection scale $R_{CC}$ is
exactly the  Jeans length, $\;R_{CC}^{-2} = g^2\rho\;$. For the current case of $\;p=\rho\;$, this is still approximately true,
\be
R_{CC}^{-2} \;=\; 2 g^2\rho\;=\;\frac{\epsilon^2}{l_s^2}\;,
\label{epsRcc}
\ee
where we have used Eqs.~(\ref{rhostring}),~(\ref{pstring}) and restored the dependence on $l_s$. Notice that, for a cosmological setup, the first equality in Eq.~(\ref{epsRcc}) implies that $R_{CC}$ is the Hubble radius $H^{-1}$ and
then $\;\epsilon\sim R^{-1}_{CC}\sim H\;$ is the Gibbons--Hawking temperature.

A different (but related) way to estimate $R_{CC}$ is to note that, if the string interactions are indeed dominated by gravity, then the total interaction energy within a causally connected region should be parametrically equal  to its gravitational energy.~\footnote{Here, we are temporarily generalizing  the  number of spatial dimensions to $\;d\geq 3\;$ to highlight the fact that the scaling of $\epsilon$ and thus the various densities are independent of this number.}
The energy of the string  within  such a region is given by
\be
E_{string}\;\simeq\;\rho R_{CC}^d\;\simeq\; \frac{\epsilon^{2}}{g^2} R_{CC}^d
\ee
and the gravitational energy can be expressed using  the Newtonian potential,
\be
E_{grav} \;\simeq\; g^2  \frac{E_{string}^2}{R^{d-2}_{CC}}
\;\simeq\; \frac{\epsilon^4}{g^2}R_{CC}^{d+2}\;,
\ee
along with the identification
$\;G_N=l_P^{d-1}=g^2 l_s^{d-1} = g^2\;$.
Then, since the interaction energy should be the same order as
the net energy of the string, it follows that
\be
\frac{\epsilon^{2}}{g^2} R_{CC}^d\;\simeq\; \frac{\epsilon^4}{g^2} R_{CC}^{d+2}\;,
\label{grav-energy}
\ee
the solution of which is, once again, $\;R_{CC}^{-2}= \frac{\epsilon^2}{l_s^2}\;$.

Yet another way to arrive at  Eq.~(\ref{epsRcc}) is to apply  a double scaling limit in which the relative deviation of the string temperature from the Hagedorn temperature is tuned  in response to the tuning of  the total length of the string $N$ \cite{strungout,emerge}.

In a ($d=3$) cosmological setup, for a yet unspecified external observer, we can interpret the above results as follows: The energy within a sphere of radius $R_{CC}$ scales as $\;E\sim R_{CC}^3\epsilon^2/g^2  \sim R_{CC}/g^2\;$ and the entropy  as $\;S\sim R_{CC}^3\epsilon/g^2  \sim R_{CC}^2/g^2 \;$.
So that, if we identify $R_{CC}$ with the inverse of the Hubble radius
$H^{-1}$ of some epoch of the Universe  and use that $g$ is the inverse of
the Planck mass $M_P$ in string units,  then the energy and entropy within a
Hubble volume go as
\be
E_H \; \sim\; \frac{M_P^2}{H}
\ee
and
\be
S_H \;\sim\; \frac{M_P^2}{H^2}\;,
\label{hubent}
\ee
which agrees with  the expected scaling relations of the energy within the horizon and  the Gibbons--Hawking entropy, respectively.  If $\;R_{CC}=H^{-1}\;$ is large (in string units),  the Hubble scale is smaller than the string scale because then $\;H\sim\epsilon M_s \ll M_s\;$.

We will be interested in situations in which not only $H^{-1}$ is large
 but  the Universe is ``large'' in another sense; namely that
its total entropy far exceeds the entropy in a single Hubble region,  $\;S_{tot}\gg S_H\;$. There will then be a large number $\;n_H=V_{tot}/H^{-3}\;$ of causally disconnected regions and  the entropy $\;S_{tot}= S_H n_H\;$  will saturate the causal entropy bound, $\;S\leq\sqrt{EV}$ \cite{ceb,hubblebound}.

Let us now discuss the symmetries of the string state. These will be essential for establishing the connection between the semiclassical perspective and the microscopic calculation of the correlation function for the perturbations. The string state is invariant in an obvious way under rotations and translations of the fiducial coordinates (except near the boundary of the state). A less obvious symmetry is one in which the fiducial coordinates are rescaled as $\;x_i \to \lambda x_i\;$, with $\lambda$ being some numerical constant. To understand the origin of this scaling symmetry, let us recall that, in a given causally connected region, the entropy of the string and so its length in string units are fixed and, therefore,  invariant under a rescaling of the fiducial coordinates. The value of $\epsilon$ or, equivalently, $R_{CC}$ is fixed  as part of the definition of the state; as  is the {\em total} entropy of the string state and, hence, the number of causally disconnected regions $n_H$. As such, all of these quantities must be  independent  of any  rescaling of the fiducial coordinates.  The same can be said of  $\rho$, $p$ and $s$, as each of these can be strictly expressed in terms of  $\epsilon$ and the coupling; see Eqs.~(\ref{consol}-\ref{pstring}).
Of course, some corrections  can be expected to lead to small breaking of these scale symmetries but not to any additional  breaking of the translation and rotation symmetries.

It seems plausible that the same reasoning requires the various quantities to be invariant under all local Weyl transformations. This is equivalent to these quantities being invariant under all of  the above transformations and, in addition, be invariant under special conformal transformations. Alternatively, it is expected on general grounds  that scale-invariant theories are conformally invariant \cite{Nakayama:2013is}. Either way,  we will  go one step further and require  the fiducial coordinates to be conformally flat.~\footnote{One might be concerned that the trace of the energy--momentum tensor is non-vanishing. However, lacking a semiclassical Einstein equation, one cannot use  properties of this tensor to constrain the symmetries of the string state.} It is unlikely that any of our conclusions are affected by this distinction.

The thermalization time scale or ``scrambling time'' \cite{scramble} of the cosmological  string state can be evaluated in a similar manner to that of BHs \cite{emerge},
\be
\tau_{scrambling} \;=\;  \frac{1}{H} \ln{\frac{S_{tot}}{S_H}}\;.
\label{tscramb}
\ee
The quantity $\ln{(S_{tot}/S_H)}$ will later be identified  as determining the number of e-folds of inflation. Equation~(\ref{tscramb}) implies that, for the total string state to reach thermal equilibrium, it  has to exist for  a long time compared to the light-crossing time of the causally connected regions;  another manifestation of the fact that the Universe needs  to be large.

A state of strings in the Hagedorn phase, for which the individual string loops
are typically long ({\em e.g.}, \cite{LT}), can undergo a phase transition from a bound state of  hot, long strings to a  state of hot radiation. This happens if the temperature  decreases by a small amount below the Hagedorn temperature, leading to a reversal in entropic dominance because of $F$ changing sign.  In which case, the stringy bound state
will decay  quickly into a gas of  small string loops; that is,
into  a radiation-dominated phase.

It is possible that such a phase transition  can be attributed to the
stringy  bound state being unstable to the emission of small loops; the
analogue of Hawking radiation. The decay time scale $\tau_{st}$ (as measured
in fiducial string coordinates) would then be the analogue  of the Page time \cite{page}, $\;\tau_{st}\sim S_H R_{CC}\;$, and the longest time scale in the problem from the internal microscopic perspective. The phase transition could also be induced by some coherent perturbation,  which might then act on a shorter time scale. For our purposes,  it is not important what brought about the phase transition, just that it occurred.

Let us now return to the potential issue of neglecting higher-order interactions. The strength of each additional string  interaction is proportional to a factor of $N/V$, the density of potential intersection points and to an additional factor of the coupling strength $g^2$. It follows that the relative strength of the
 $(n+2)$-string interactions in comparison to the strength  of the 2-string interactions is proportional to
\be
\left(g^{2}s\right)^n\;=\; \epsilon^{n}\sim\left(\frac{l_s}{R_{CC}}\right)^n\;,
\ee
where we have again used $\;s=N/V\;$ and  Eq.~(\ref{consol}).
One can see that these are
actually $\alpha^\prime$ corrections,
being proportional to powers of $l_s/R_{CC}$.  Similarly, higher-order string-loop corrections are proportional to higher powers of $\;g^2\;$.
Further recalling that the free energy is expressible  as a power series in
$\;s=\epsilon/g^2\;$,  one can see that the higher-order interactions are thus suppressed provided that
$\;\epsilon\ll g^2 \ll 1\;$.~\footnote{At a first glance, $\epsilon$ and $g$
look to be the same order because the Friedmann equation  $\;H^2\sim g^2\rho\;$
with $\;\rho\sim M_s^4\;$ and $\;H\sim \epsilon\;$ implies
that $\;\epsilon^2 \sim g^2\;$. However, when one takes  numerical values seriously, it becomes clear that $H^2$ is suppressed significantly below
this estimate; see Subsection~3.1.}

\subsection{Semiclassical geometric perspective: flat slicing of de Sitter space}

As discussed in Section~1, in looking for a semiclassical geometric description of the initial state,  if one is  willing to ignore the entropy and just focus on the state's mechanical aspects, then the conclusion would be  that $\;p=-\rho\;$. This follows directly from the thermodynamic identity $\;p+\rho = T s\;$  given that $\;s=0\;$. The resulting state can then  be described by a time-independent geometry within
the framework of general relativity.

Let us begin here  by specifying  the $\;p=-\rho\;$  geometry for a spherical volume of radius $L$.
It is described by the time-independent line element
\be
ds^2\; =\; -f(R) dT^2 + \frac{1}{{\widetilde f}(R)} dR^2 + R^2 (d\theta^2+ sin^2 \theta d\phi^2)\;,
\ee
along with an  energy--momentum tensor of the form
$\;T^\mu_{~\nu}= - \rho~ \delta^\mu_{~\nu}\;$.
The solution of the Einstein equations then works as
\be
ds_{SP}^2\; =\; -(1-\frac{R^2}{L^2}) dT^2 + \frac{1}{(1-\frac{R^2}{L^2})} dR^2 + R^2 (d\theta^2+ sin^2 \theta d\phi^2)\;,
\ee
for which
\be
\frac{1}{L^2}\;=\; \frac{8 \pi G}{3} \rho\;.
\ee
and $\;R/L \le 1\;$.
This solution describes the static patch of dS space,  which corresponds to the interior of the cosmological horizon at $\;R=L\;$. The energy density $\rho$ from this perspective is attributed to a cosmological constant $\;\Lambda =3/L^2\;$. Notice that the geometry is spatially flat; one cannot introduce a source of spatial curvature  without violating  the equation of state $\;p=-\rho\;$.

If one wishes to formally extend this geometry to regions beyond the horizon,
an extra leap of faith has to be taken. The formal extension to  a Universe containing a large number of causally disconnected static patches --- the geometry of an inflating Universe ---   is achieved by transforming  to planar dS coordinates $r$, $t$ via
$\;R = e^{H t} r\;$ and $\;e^{-2Ht}= e^{-2HT} -R^2H^2\;$ with $\;H=1/L\;$. The line element then adopts the familiar  form
\be
ds_{planar}^2\; =\; - dt^2 + a(t)^2 \sum_i dx_i^2 \;,
\ee
with $\;a(t) = e^{H t}\;$ and $\;r^2=\sum_i x^2\;$.
It should be kept in mind that the dS spacetime, whether in its static or planar form, is an alternative ``dual" description  of the initial stringy state. The string fluid is not the source in the Einstein equations that leads to a dS solution.

In spite of the apparent time dependence in the planar description, the geometry is still inherently time independent as the static-patch metric makes clear.  Real  time dependence, from this perspective, occurs only when the geometry is modified and some physical clock is added. It follows that the decay of the initial dS spacetime into one filled with  hot radiation ---  which from the perspective of the late-time FRW observer  corresponds to the end of inflation ---  has to be described by adding some external, time-dependent modification to the dS geometry. Moreover, meaningful observations can only be made by the FRW observer after the state has
further  decayed  through the properties of the matter. This will be discussed
in the  next subsection.

The isometries of dS space are important for fixing the structure of correlation functions of perturbations (see, {\em e.g.},  \cite{Maldacena:2011nz,Bzowski:2013sza,Kundu:2014gxa,nima}) and will also be  important here  for establishing the connection between the string state and
its effective  dS description. These are essentially the rotations and boosts of a hyperboloid in a flat spacetime with one additional dimension.
When translated  to planar coordinates,
these include  the obvious  translations and rotations, along with   dilatations (or scalings) and  special conformal transformations. In other words, the conformal group.

\subsection{Comparing the microscopic and the effective semiclassical perspectives}

From the microscopic  perspective,  the phase transition from the initial state of strings into a state of  hot radiation  marks the end of inflation. Moreover, this initial state  is known and sets the boundary conditions for the semiclassical evolution of gravity and matter from that point on. The matter content is also known in principle as well as the temperature, pressure, energy and entropy densities. Alternatively, according to the semiclassical perspective, the inflationary phase of the  Universe is described by an  empty, time-independent half of dS space which is covered by the planar coordinates. Once inflation is over, this  distinction is inconsequential.

But, as discussed in Section~1,  the  score keeper  for the state of the Universe is  the FRW observer. This observer uses the same comoving  coordinates $r$ and $t$ as for the planar dS space  but can only make observations   after inflation has finally  ended. Meaning  that the end time of inflation is the analogue of the Schwarzschild radius in the BH  case and the inflationary past is the analogue  of the BH interior.  In short, the status of the FRW observer after inflation is quite similar to  that  of an observer who remains outside of a BH. It is after  the end of inflation that the FRW observer
can measure  the average temperature, energy density, temperature anisotropies, {\em etc}. The measurements could be done, for example, now or when the radiation of the cosmic microwave background (CMB) decoupled from other matter (or even at  earlier times in principle).

\begin{figure}[t]
  \centering
  \vspace{-.5in}
  \includegraphics[width=0.75\textwidth]{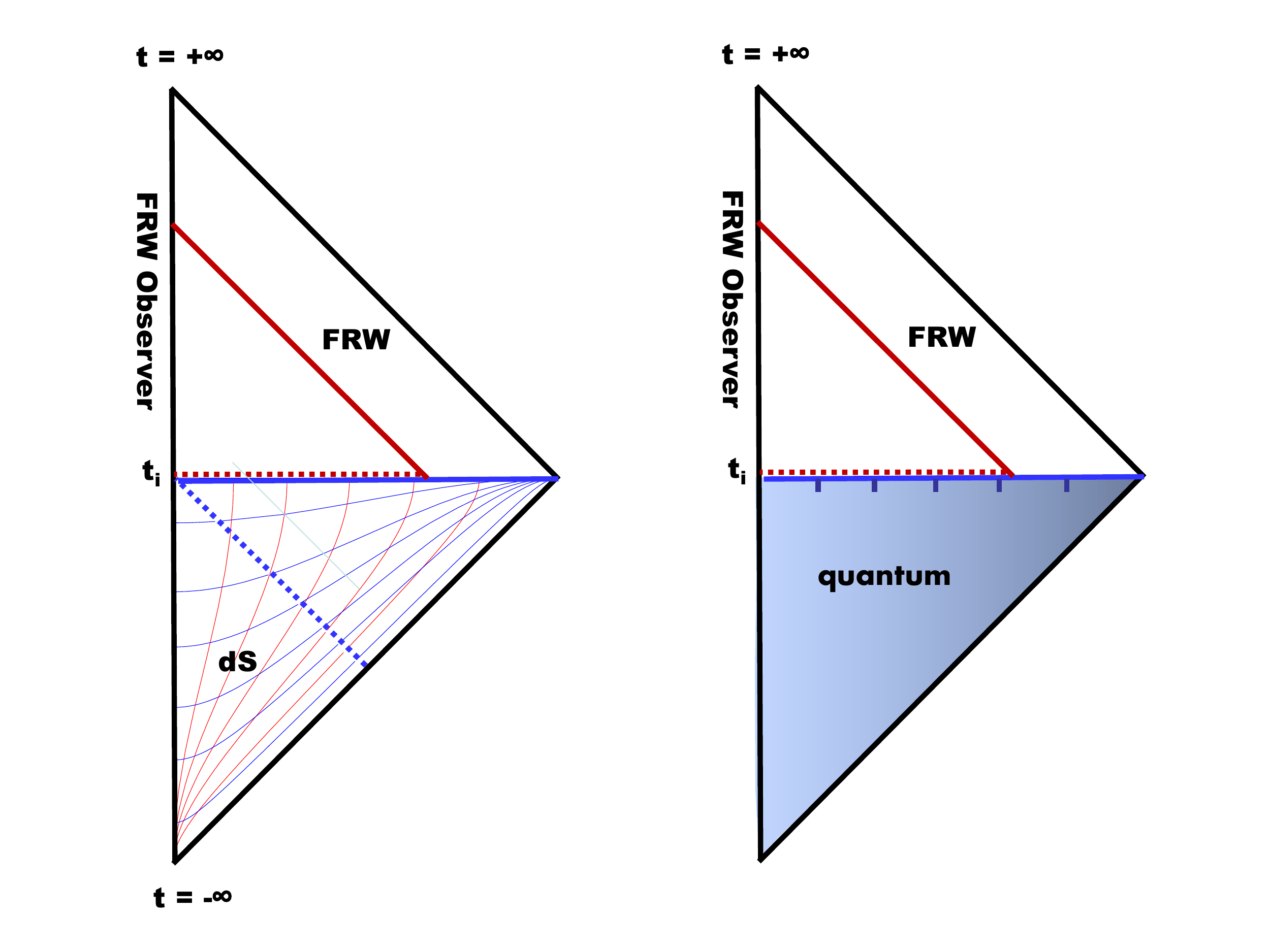}
  \caption{Universe as viewed by the FRW observer. Shown are Penrose diagrams depicting the semiclassical interpretation (left) and the microscopic interpretation (right). Also displayed are the FRW observer's past light cone and the number of causally disconnected regions that he can see. In the lower half of the semiclassical diagram, lines of constant planar coordinates $t$ (approximately vertical) and $r$ (approximately horizontal)  are shown.
  }\label{Fig:penrose1}
\end{figure}

From the FRW observer's perspective,  the Universe starts its evolution in a thermal state, and so he would  need to  invent a  pre-history to explain  the observable Universe. He can do this in the same way that a semiclassical observer invents a description of the BH interior \cite{BHfollies}. The FRW observer
will  conclude that the Universe was exponentially expanding at some epoch in its pre-thermal history, and  he would then  need to explain to himself how this strange era in the early Universe came about.   The inflationary paradigm provides just such an  explanation. Accepting this ``fable'', the FRW observer
also has to explain the decay by  smoothly matching the initial dS phase to the FRW phase. Matching in this way, he concludes that the dS phase could not really be time independent and that the  ``clock''  keeping  track of this (faux) evolution is the inflaton field. From this point of view, the inflaton does not need to correspond to any physically real field (or combination of fields) that exists in the FRW phase.

Let us reemphasize our essential  point that the inflationary paradigm is an invented effective history of the Universe. What is physical and real are the results of the measurements that are made by an FRW observer after the end of inflation ({\em i.e.}, after
the time when the whole Universe is ``created'');  in particular, the measurements of the temperature anisotropies.~\footnote{Note, though, that the FRW observer can only   measure  a portion  of
the post-inflationary Universe at finite times.  However, the whole state does  become  observable as $t$ increases.}  If these can be predicted correctly, this is all there is to it. The rest of the theory consists of conceptualizations that can never be verified and must forever
remain  ambiguous.

A remarkable fact is that the symmetries and causal structure  of the string state and dS space are  very similar. The string state is invariant, at the very least,  under rotations, translations and scalings (or dilatations), as is its causal connection scale $R_{CC}$, which then sets the size of the horizon $1/H$. Meanwhile, in the case of dS space, the isometries are those of  the conformal group which additionally contains   special conformal transformations. As explained earlier, it is likely that the string state is also invariant under the full conformal group and we proceed, for the sake of simplicity,
with this as our premise.  This commonality will be important for determining the correlation functions of the perturbations and showing that both perspectives lead to the same scale-invariant spectrum.

A comparison of the two alternative histories that the FRW observer can choose from is shown in Fig.~\ref{Fig:penrose1}.  The semiclassical inflationary paradigm and the stringy initial state lead to the same observational results if the parameters of the inflationary model are chosen appropriately.

\section{Inflation}

As argued earlier, any invented pre-history of the Universe that is self-consistent, agrees with observations and obeys the laws of physics is as good as  any other. The advantage of our microscopic description is that it
unambiguously determines the parameters in the effective description of inflation  and, as a by-product, resolves some of the issues that were discussed in the
introductory section.  Let us next explain how the effective parameters are determined. We do  not discuss precise quantitative predictions here but, rather,
elaborate on  some of the qualitative features. A more detailed comparison to the  observations will be deferred to a future article \cite{futureCMB}.

\subsection{Scale of inflation}

The energy density of the radiation just after the Hagedorn phase transition is $T_{Hag}^4$ and recall that  the Hagedorn temperature relates   to the string mass as $\;T_{Hag}=\frac{M_s}{4\pi}\;$. Since the Hubble parameter  is given by the Friedmann equation,
\be
3 H^2\; =\; \frac{1}{m_P^2} T_{Hag}^4 \;,
\label{friedman}
\ee
with $m_P$ denoting  the reduced Planck mass  $\;m_P^2 = 1/(8 \pi G)=M_P^2/(8\pi)\;$, it follows that
\be
H \;=\; \frac{1}{\sqrt{3} (4\pi)^2} \frac{M_s^2}{m_P} \;\simeq\;
10^{14}~{\rm GeV} \left(\frac{M_s}{2.4\times 10^{17} GeV}\right)^2\;.
\label{hubblybubbly}
\ee
We have kept track of numerical factors, both  here and below,  because their large values are important for ensuring the viability of the model.

In our model is that the scale of inflation $H$ is expressed in terms of the fundamental string scale $M_s$, which  is expected to be somewhat below the Planck scale $\;M_P=1.2\times 10^{19}\;$~GeV. In \cite{Brustein:2002mp},
for instance,  $M_s$ was estimated to be of the order of $10^{17}$~GeV for several types of string theories. In other scenarios it can, however, be significantly lower or, sometimes, even somewhat higher.

The entropy within a Hubble volume $S_H$ at the beginning of the radiation era is given by the entropy of the string state
({\em cf}, Eq.~(\ref{hubent})),
\be
S_H \;=\; \pi  \frac{M_P^2}{H^2} \;\simeq\; 10^{11} \;\left(\frac{2.4\times 10^{17}\;{\rm GeV}}{M_s}\right)^4\;,
\label{initialS}
\ee
where Eq.~(\ref{hubblybubbly}) has been used in the right-most relation.
This large entropy will later be related to the strength of the curvature perturbations.

Unlike for the entropies, the temperature of the radiation $T_{rad}$ is parametrically  lower than the Hagedorn (string) temperature. This is because of the large number of constituent particle species, $\;N_{species}\gg 1\;$ that are expected to be  in thermal equilibrium at this high
a temperature, so that $\;T_{Hag}^4 = N_{species} T_{rad}^4\;$.

\subsection{Duration of inflation}

The next task is to determine the duration of inflation $\tau_{inflation}$, which is a free parameter in the effective description of inflation. The accepted constraint on $\tau_{inflation}$ is that it should be long enough to explain the entropy and flatness of the observed Universe; it could, however, be much longer. Here, we will use the entropy as the physical quantity that determines the duration of inflation in the microscopic description.

Let us recall that the entropy of the stringy bound  state is the total length of string in string units. We are assuming (but will later relax our assumption) that this string entropy is fully converted into  radiation entropy  at the phase transition which marks the ends inflation. The number of e-folds that the FRW observer has to postulate is, from his perspective,  determined by the increase in volume  that is required to explain the difference between the initial entropy $S_H$ in Eq.~(\ref{initialS}) and the total entropy of the Universe $\;S_{tot}=n_HS_H\;$,
\be
\label{boblawblaw}
e^{3 N_{e-folds}}\;=\;\frac{V_{tot}}{H^{-3}}\;=\;n_H\;=\; \frac{S_{tot}}{S_H}\;.
\ee

Further  recall that $n_H$ is the number of causally disconnected patches which
 the string state encompasses from an external semiclassical  perspective.
Since the entropy is fully converted into radiation, the amount of inflation is minimal; just enough e-folds as would be necessary to ``explain'' the large number of causally disconnected patches. In this way,  it is the value of $S_{tot}$ which determines the  number of e-folds that the FRW observer has to postulate,
\be
N_{e-folds}\;\leq\;\frac{1}{3} \ln{\frac{S_{tot}}{S_H}}\;.
\ee
The right-hand side sets  an upper bound because, if
there were any  additional entropy-generating mechanisms  after inflation, it would reduce the number of e-folds that are required to explain $S_{tot}$.

In our Universe $\;S_{tot}\sim 10^{88}\;$  so that, if there are no additional sources of entropy, the requisite number of e-folds should be
\be
N_{e-folds}\;\sim \;\frac{1}{3} \ln{\left(\frac{S_{tot}}{S_H}\right)} \;\simeq\; \frac{1}{3} \ln 10^{76} \;\sim\; 60\;.
\ee
If we convert this number of e-folds into a dimensional  duration of inflation, then
\be
\tau_{inflation}\;=\;H^{-1} N_{e-folds}\;\sim\;l_s (g^2 S_H)^{1/2} \ln{\left(\frac{S_{tot}}{S_H}\right)}\;.
\ee
Notice that the duration of inflation can be  expressed strictly in terms of the parameters of the string state.
The resulting time scale is, remarkably, of the same  order as the scrambling time of the string state in Eq.~(\ref{tscramb}), $\;\tau_{inflation}\sim \tau_{scrambling}\;$.

In spite of this last observation, let us emphasize that the number of e-folds is not directly related to any specific time dependence.  It is, rather,  part of the story that an FRW observer needs to tell in order to explain the fact that the initial entropy is much smaller than the final one. In addition, the FRW observer needs to introduce this artificial notion of  time dependence to explain why inflation ends and how perturbations originate.

The above discussion highlights the fact that the maximal entropy state does not solve the so-called homogeneity (or smoothness or size) problem in spite of
its resolution of  the flatness problem.  A solution of  the homogeneity problem would amount to explaining why the stringy state extends over many horizon regions or, alternatively, why it is so long lived. As mentioned in the Introduction, it is likely that one needs a long phase of contraction (whether it be ultra-slow \cite{ekpyrotic} or accelerated \cite{pbb}) to explain the desired
degree of smoothness. However, in the context of our discussion, this is a topic that is not directly relevant to the subsequent evolution of the post-inflationary Universe.

\subsection{Perturbations}

Being strongly quantum in its nature, the initial stringy state of  maximal entropy has to be fluctuating quantum mechanically. The objective here is to understand how these perturbations can be compatible with cosmological observations.

The essential properties of the perturbations in inflation are that they ``freeze" on scales larger than the horizon and are ``scale invariant''. The meaning of ``freezing" is that their amplitudes no longer oscillate as a function of time but are (approximately) constant. The proper wavenumber $p=|\vec{p}|$ of the perturbations when they freeze is always the horizon scale, $\;p=H\;$,  but  the perturbations are deemed  as scale invariant when their amplitudes are independent of their comoving wavenumber $\;k=ap\;$ at the time of freezing. A simple  way to understand both properties is to trade cosmological time $t$ for conformal time $\;\eta =\int \frac{dt}{a}$ so that $\;\eta=-1/Ha\;$. Then, the
horizon-crossing condition for $p$ becomes $\;|\eta k| =1\;$. Hence, after crossing,  the oscillating part of the perturbations in comoving coordinates $\;\delta\sim e^{ik\eta};$ depends on neither time nor wavenumber.

We will eventually reveal the origins of these two properties  from the microscopic perspective while determining the amplitude of the perturbations in terms of the string scale. But even without a detailed analysis, one can anticipate these features simply because the initial string state shares with dS space the critical  property of invariance under scaling transformations.
Moreover, the change in the nature of the stringy perturbations across the horizon scale was already implicitly discussed in our description of
 the calculation of $R_{CC}$  and  is especially apparent from Eq.~(\ref{perteq}).
 The nature of these perturbations is described in detail in \cite{duality}; in particular, they are shown there to be  approximately constant ``outside the horizon" when $\;k^2 < R_{CC}^{-2}\;$.  This constancy for perturbations  that have escaped from their Hubble-sized domain  may  seem strange  insofar as there are $n_H$ independent domains and, as such, variances should be suppressed by a factor of $1/n_H$. However, from a quantum-mechanical perspective, it is natural to expect that  the  patches are not so independent but, rather, correlated through quantum entanglement. And, as is well known, quantum correlations persist over an arbitrarily long length scales.

A simple way to quantify the strength of the perturbations in the stringy description is by considering the entropy fluctuations  in some region. To this end, let us consider some observable quantity $O$ in a spherical region of radius $R$ (in terms of the stringy fiducial coordinates). Then the relative strength of the quantum fluctuations  of $O$ in the specified region can be defined as  $\;\delta^2_O(R)\equiv \langle\delta O^2\rangle_R/\langle O(R)\rangle^2\;$. Our main focus is on  perturbations on the scale of the horizon $\;H \simeq 1/R_{CC}\;$, but we will later comment on longer length scales.

Let us next recall how the fluctuation strengths are evaluated from the  free energy
of the string state \cite{emerge}.
The essential point here is that, at fixed values of  ``temperature''  $\epsilon$, the entropy density $s$ is essentially conjugate to the volume $V$. To see this,
one can  rewrite Eq.~(\ref{FES2}) as
\be
-\frac{F}{T_{Hag}\epsilon}\;= \; s V - \frac{1}{2\epsilon} g^2 s^2 V \;.
\label{FES3}
\ee
The variance of the volume fluctuations at fixed temperature  can now be evaluated
 in the standard way,
\be
(\delta V)_{\epsilon}^2 \;=\; \frac{\partial^2 (F/T_{Hag}\epsilon)}{\partial s^2}
 \;=\; \frac { g^2 V}{\epsilon} \;.
\ee
Substituting for  $\epsilon$ from   Eq.~(\ref{consol}) and dividing
by $V^2$,  we  now  obtain
 \be
\delta^2_V (R) \;=\; \frac{1}{S(R)}\;,
\ee
for some region of size $R$.

The conjugacy  of $s$ and $V$ ensures that their respective  fluctuations have the  same relative strength, so that
\be
\delta^2_s(R_{CC})\;\simeq\; \frac{1}{S(R_{CC})}\;.
\label{deltaSH}
\ee
And, since $s$ and $\rho$ differ only by a single factor of  (constant) $\epsilon$,
\be
\delta^2_{\rho}(R_{CC})\;\simeq\;\frac{1}{S(R_{CC})}=\frac{1}{\pi} \frac{R^2_{CC}}{M_P^2}\;
\label{deltaEH}
\ee
or, in more conventional  notation,
\be
\delta^2_\rho(H)\;\simeq\;\frac{1}{\pi} \frac{H^2}{M_P^2}\;,
\label{deltarhoH}
\ee
where
the argument $H$ has been used  as short hand for a
Hubble-sized length scale  $\;R_{CC} \simeq 1/H\;$.
The standard inflationary result scales in the same way, as
$\;P_{\rho}(k)=\frac{2}{\pi} \frac{H^2}{M_P^2}\;$ is the
closely related  power spectrum for the tensor perturbations \cite{mukhanov}. As already mentioned, we will conduct a precise quantitative comparison between our results and those of the inflationary paradigm in a separate publication \cite{futureCMB}.

From the semiclassical perspective, Eq.~(\ref{deltarhoH}) quantifies the strength of the tensor curvature perturbations rather than the scalar curvature perturbations, which will be discussed next. These tensor perturbations are, therefore, the most direct link between the stringy description and the effective model of inflation. Using either  Eq.~(\ref{hubblybubbly}) or~(\ref{initialS}), one can estimate their strength in terms of the value of the string scale  $M_s$.

A  less direct link is provided by the value of the gauge-invariant scalar curvature perturbation $\zeta$ \cite{mukhanov}. We will rely on the relationship between $\zeta$ and the perturbation in the number of e-folds $\delta N_{e-folds}$, as used in the ``separate Universe" approach and the $\delta N$ formalism  \cite{separate1,separate2} for the calculation of super-horizon perturbations,
\be
\zeta \;=\; \delta N_{e-folds}\;.
\ee
This identity will be used here  as the defining relation of the gauge-invariant scalar perturbation in the maximal entropy state,
\be
{\zeta}_{st}\; \equiv\; \delta N_{e-folds}\;.
\label{zetast1}
\ee
As shown below, the value of $\delta N_{e-folds}$ can be expressed in terms of quantitites that are well defined for the maximal-entropy state.

The number of e-folds in the maximal entropy state can be expressed as in Eq.~(\ref{boblawblaw}),
\be
N_{e-folds}\;=\; \int dt \;\frac{1}{3} \partial_t \ln V \;\simeq \;\int dt \; \partial_t \ln S\;,
\label{e-folds1}
\ee
from which it follows that
\be
\delta N_{e-folds}\;\simeq\; \int dt\;  \partial_t \frac{\delta S_H}{S_H}\; \simeq\; \int dt \;  H \left.\frac{\delta s}{s}\right|_H\;.
\label{e-folds2}
\ee
It can  then be concluded that
\be
\zeta_{st} \;=\; \int dt \; H \left.\frac{\delta s}{s}\right|_H\;.
\label{zetast2}
\ee
It is also noteworthy  that
\be
\dot{\zeta}_{st}\;=\; H  \left.\frac{\delta s}{s}\right|_H\;,
\ee
similar to the case in which the perturbations are non-adiabatic \cite{mukhanov,zetaCurvaton}.

It should be emphasized that  the above time derivatives refer to the time coordinate $t_{st}$ of the  string state's fiducial (conformally flat) system and not that of the  planar dS space. However, as both descriptions share the property of conformal invariance, we have the freedom to match their respective coordinate systems with one another and with those of the FRW observer at the surface $\;r=H^{-1}\;$ (equivalently, at  $\;r_{st}=R_{CC}$).   In other words, once at or  outside the horizon,  all relevant observers can be assigned a common definition of  time, and we can subsequently adopt the coordinate  $t$ of planar dS  without any loss of generality.   Nonetheless, as explained  below (also see \cite{nima}), any  time dependence in the calculation is something of a red herring and so one's particular choice of time coordinate is almost besides the point.

The magnitude of the scalar curvature perturbation
at the horizon can be evaluated as follows:
\be
\langle \zeta_{st}^2 \rangle_{H}\; =\; \int dt'\int dt'' H^2 \langle
\frac{\delta s}{s}(t')\frac{\delta s}{s}(t'') \rangle_{H}
\;\simeq\; H\int dt \delta^2_s(H)
\;\simeq\;  H\int dt \delta_\rho^2({H})  \;.
\label{zeta2}
\ee
Here, $\langle \frac{\delta s}{s}(t')\frac{\delta s}{s}(t'') \rangle_{H}\sim  H^{-1}\delta(t'-t'')\;$ has been used. The reason that a delta function appears here is
basically the same reason that it appears in the standard inflationary calculations. The horizon-crossing constraint --- or its casual-connection analogue   --- means  that at times such that $\;H|t'-t''|>1\;$, the perturbations $\zeta_{st}(t')$ and $\zeta_{st}(t'')$ are uncorrelated. From Eq.~(\ref{zeta2}), one can observe that $\zeta_{st}$ evolves outside the horizon, again, as in the case of non-adiabatic perturbations.

The remaining integration over time can be interpreted as the time to probe
the {\em total} length of string, which amounts to an integration over
the number $n_H$ of independent domains.  To see this, one can apply
 $\;Hdt\simeq dN_{e-folds} \;$ to obtain
\be
\langle \zeta_{st}^2 \rangle_{H}
\;\sim \; \int dN_{e-fold} \delta_\rho^2({H}) \;\sim\; \int
d\ln{n_H}\;\delta_\rho^2({H})\;\;,
\label{scalarR}
\ee
where Eq.~(\ref{boblawblaw}) has been used.
The last  equality also makes it clear that Eq.~(\ref{scalarR}) is  expressed entirely in terms of the microscopic description.

If one assumes that $\delta^2_\rho({H})$ remains  approximately constant over  the $n_H$  disconnected regions, the above integral simplifies to
\be
\langle \zeta_{st}^2 \rangle_{H} \;\sim\;  N_{e-folds} \delta^2_\rho({H})\;.
\ee
The factor of $N_{e-folds}$  represents an enhancement factor of the scalar perturbation over the tensor perturbations. Indeed, the tensor-to-scalar ratio $r$ is given by
\be
r \;=\;\frac{\delta^2_\rho(H)}{\langle \zeta_{st}^2\rangle_{H}}\;\simeq\; \frac{1}{N_{e-folds}}\;.
\ee
The calculation of the scalar perturbations and the resulting enhancement does not depend explicitly on any  deviations from scale invariance. The quantity $\zeta_{st}$ therefore shares the properties of the standard gauge-invariant scalar perturbation $\zeta$ --- it is the physical clock.

The  calculation of the  $\zeta_{st}$ correlation function can, as alluded to, also be done in a way that eliminates the time dependence altogether. Let us reconsider the double time integral in Eq.~(\ref{zeta2}) and rewrite each integral in terms of conformal time, $\;\int dt \to \int a d\eta\;$.  Applying the derivative of the comoving form of horizon-crossing constraint $\;kd\eta+\eta dk=0\;$, one then has
$\;\int  a d\eta \to -\int \frac{\eta}{k} a dk  = \int\frac{1}{kH} dk\;$.
If one also approximates the time derivatives by $H$, the double time integral in Eq.~(\ref{zeta2}) becomes a double integral in $k$-space,
\be
  \langle \zeta_{st}^2 \rangle_{H}\; =\; \int d(\ln k')\int d(\ln k'') \langle
\zeta_{st}(k')\zeta_{st}(k'') \rangle_{H}\;.
\ee
Further imposing the delta function resulting from momentum conservation, we obtain
\be
\langle \zeta_{st}^2 \rangle_{H}\; \sim\;\int d(\ln k)\langle \zeta_{st}^2(k)\rangle_H \;\simeq\; \int d(\ln k)P_{\zeta_{st}}(k)\;,
\ee
where the right-most equality follows from the standard relation
between a two-point correlation function in position space and its associated  power spectrum. From Eq.~(\ref{zeta2}), it can then be deduced that
\be
\langle \zeta_{st}^2 \rangle_{H}\;\simeq\; N_{e-folds}\int d(\ln k)P_{\rho}(k)\;.
\ee
This makes clear that both the time-dependence of the scalar perturbations and their scale invariance can be presented in a way that can be viewed purely in terms of the Fourier modes of the string state.

By  working within the microscopic picture,  we did not need to introduce nor rely on  any time dependence. Notice also that the enhancement of the scalar perturbations with respect to the tensor perturbations was arrived at in a way that does not depend explicitly on any deviations from a scale invariant state.  The tensor fluctuations, which represent the fundamental quantity from this perspective, depend only on the thermodynamic properties of the initial string state. Meanwhile, the enhancement factor for the scalar modes can be attributed to the largeness of the initial state rather than a contrived period of dynamical
evolution.  Additional deviations could arise from sub-leading terms and departures from exact equilibrium; see below.

Now what happens on scales larger than the horizon? The answer is simple:
essentially the same as in the standard inflationary picture. This
is because both the freezing of the amplitudes (or their growth) and the
scale invariance of the spectrum are direct consequences of the scaling
symmetry of the dS space and the form of the perturbation equation. As we
have shown,  the corresponding  symmetries and perturbation equations of
the  microscopic string state have the same relevant properties as those in dS space.  The scaling symmetry ensures, in particular, that the position of the horizon scale
and the causal-connection scale $R_{CC}$ are  invariant features of dS space
and the microscopic string state, respectively.

More formally and as reviewed recently in \cite{nima,toniggi},
given a two-point function $\langle \delta\phi^2\rangle$ in position space that is invariant under dilatations $x^a\frac{d}{dx_a}$,  its two-point function
in Fourier space ({\em i.e.}, its power spectrum) has to have the following form at {\em all} trans-Hubble scales:
\be
P_{\phi}(k)\;=\;\frac{k^3}{2\pi^2}\langle\delta\phi(\vec{k})\delta\phi(-\vec{k})\rangle \;=\; C_{\phi}(H)\;,
\ee
where $C_{\phi}$ is a function of $H$ only.
As the tensor and scalar perturbations inherit the same set of symmetries from the string state as they do from dS space, all
that is left is to fix the value of  the $C{\rm 's}$, which we know
to be $\;C_{\rho}(H)\simeq H^2/M_P^2\;$ from Eq.~(\ref{deltarhoH})
and then  $\;C_{\zeta_{st}}(H) \simeq N_{e-folds} H^2/M_P^2\;$
due to the enhancement.

Let us now discuss possible deviations from Gaussianity and scale invariance.
In the approximation that the  free energy of the string state is quadratic and the state is in equilibrium, the fluctuations of the various quantities are strictly Gaussian.

Deviations from Gaussianity then depend on the relative strength of  the higher-order string interactions.
As discussed in Section~2, the relative strength of the  $(n+2)$-string interactions
in comparison to the strength of the 2-string interactions scales as
$
\left(\frac{g^{2}N}{V}\right)^n\;=\; \epsilon^{n}\;=\;\left(\frac{H}{M_s}\right)^n\;.
$
These are, as
already mentioned,  highly suppressed   $\alpha^\prime$ corrections,  which indicates that the perturbations are approximately Gaussian.   Higher-order string-loop corrections (and combinations thereof)  are also possible but these would
of course  be  suppressed by additional factors of
$\;g^2\;$.
Deviations from strict scale invariance are more subtle and will be discussed in more detail in \cite{futureCMB}.

\section{Summary and Conclusion}

We have proposed a microscopic model of inflation and
were able to show that, qualitatively, all the essential features of inflation are reproduced.
Our model is premised on a simple idea, that the resolution of a spacelike
singularity --- whether it be at the center of a BH or as a precursor to
the big bang --- requires large deviations from semiclassical physics over
horizon-sized scales. Large deviations in this sense means that the
region of spacetime lacks a description in terms of a semiclassical geometry.
Such a situation occurs for strongly coupled states of matter, which are
synonymous with states of extremely large entropy. The Hagedorn phase
of closed strings is one notable example  of just such a state.

From our perspective, any description of the early
Universe up to the end of the inflationary phase is as valid as any other, provided
that it is self-consistent, agrees with observations and obeys the known laws of physics.
This puts the late-time or FRW observer on equal footing with the exterior
observer in the BH case, as both can only ``make up  stories'' about the
early Universe and the BH interior respectively. Our story is one of  a  cosmological  picture that is devoid of singularities but deprived of a semiclassical geometry at early times.
 On the other hand, the  observer who is determined to maintain a semiclassical description of the inflationary period will take our picture from one extreme, maximal entropy and maximally positive pressure, to another, zero entropy and maximally negative pressure. The later extreme being the more conventional dS description of inflation.

What is then to be gained from  our stringy description of
the early Universe? The answer being that our model  fixes the parameters
of inflation in terms of two fundamental quantities, the string mass and
the string coupling. As an added bonus, this model can be connected to
BH physics, for which  the same two parameters can also be probed.
The only other input is  the minimal number of e-folds that is needed to explain the size of the Universe. Our description also manages to evade some
of the usual conceptual issues haunting inflation, such as the identity
of the inflaton and the self-consistency problems, such as eternal inflation, that ensue from
that  perspective. Furthermore,  the  inflaton models tend to have many adjustable parameters and a myriad of moving parts. In short, the lack of opportunities for ``tweaking''   means that  our  framework is predictive and will be much easier to substantiate (or falsify). The relatively small suppression of the tensor perturbations with respect to the scalar perturbations suggests that they could be discovered sooner than later.   A sequel to the current article which includes a substantially  more quantitative analysis of the observable consequences of our model will be presented in due course \cite{futureCMB}.

\section*{Acknowledgments}

We would like to thank  Marko Simonovic, Paul Steinhardt and Gabriele Veneziano for discussions and in particular Toni Riotto, for fruitful discussions and useful suggestions. The research of RB   was supported by the Israel Science Foundation grant no. 1294/16. The research of AJMM received support from
an NRF Incentive Funding Grant 85353 and a Rhodes University
Discretionary Grant RD51/2018. AJMM thanks Ben Gurion University for their  hospitality during his visit.

\end{document}